\documentclass[aps,pre,a4paper,10pt,twocolumn,showkeys,reprint,floatfix]{revtex4-1}
\usepackage[T1]{fontenc}
\usepackage{amsmath}
\usepackage{amssymb}
\usepackage{bm}
\usepackage{graphicx}
\usepackage{csquotes}
\usepackage[colorlinks=true,hyperfootnotes=true,breaklinks=true]{hyperref}
\usepackage{color,soul}
\usepackage{psfrag}
\usepackage{caption}
\usepackage{subcaption}

\setstcolor{red}
\setulcolor{green}

\begin{document}

\title{Cleaning large-dimensional covariance matrices for correlated samples}

\author{Zdzisław~Burda}
\email{zdzislaw.burda@agh.edu.pl}
\thanks{\includegraphics[scale=0.1]{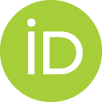}~\href{https://orcid.org/0000-0002-9656-9570}{0000-0002-9656-9570}}
\affiliation{\href{http://www.agh.edu.pl/}{AGH University of Science and Technology, Faculty of Physics and Applied Computer Science, al. Mickiewicza 30, 30-059 Kraków, Poland}}

\author{Andrzej~Jarosz}
\email{jedrekjarosz@gmail.com}
\thanks{corresponding author, \includegraphics[scale=0.1]{ORCID.png}~\href{https://orcid.org/0000-0002-5857-9241}{0000-0002-5857-9241}}
\affiliation{\href{https://www.crisil.com/}{CRISIL Irevna Poland Sp z o.o., ul. Świętego Mikołaja 7, 50-529 Wrocław, Poland}}

\keywords{random matrix theory, Mar\v{c}enko-Pastur equation, free probability, shrinkage estimators}

\begin{abstract}
We elucidate the problem of estimating large-dimensional covariance matrices in the presence of correlations between samples. To this end, we generalize the Mar\v{c}enko-Pastur equation and the Ledoit-P\'{e}ch\'{e} shrinkage estimator using methods of random matrix theory and free probability. We develop an efficient algorithm that implements the corresponding analytic formulas, based on the Ledoit-Wolf kernel estimation technique. We also provide an associated open-source Python library, called \texttt{shrinkage}, with a user-friendly API to assist in practical tasks of estimation of large covariance matrices. We present an example of its usage for synthetic data generated according to exponentially-decaying auto-correlations.
\end{abstract}

\maketitle

\section{Introduction}

In experimental research, when working with large datasets, one often faces a generic problem of determining correlations between multiple entities of interest, on the basis of observed data. This problem, in its minimalistic form, can be formulated as follows: One performs $T$ measurements of a statistical system with $N$ degrees of freedom, $Y_{i}$, $i=1, \ldots, N$, and collects the observations in a matrix $Y = [y_{i t}]$ of dimensions $N \times T$, where $y_{i t}$ is the $t$-th measured value of the $i$-th entity. The challenge lies in estimating from this dataset the underlying two-point covariances,
\begin{equation}\label{eq:C}
    C_{i j} = \langle Y_{i} Y_{j} \rangle ,
\end{equation}
collected in the so-called \textit{population} (a.k.a. \textit{true}, or \textit{signal}) \textit{covariance matrix}, $C = [C_{i j}]$. (We assume throughout the article that the data has already been centered, $\langle Y_{i} \rangle = 0$.)

This setting is very general, commonly encountered in physics, finance, genomics and bioinformatics, signal processing and acoustics, image recognition, speech recognition, cancer research, climatology, neuroscience, and many other areas. Below we sketch several such occurrences, also in each case providing relevant orders of magnitude (marked with the symbol $\sim$) for $N$ and $T$; we will see that, crucially, both these numbers are typically quite large and of comparable value.

For instance, in lattice QCD, one may consider \cite{RSWBSW} neutron-antineutron two- and three-point correlation functions, described using $N$ ($\sim$ a few dozen) parameters related to the number of dynamics time steps. One simulates field configurations on the lattice, obtaining thereby $T$ samples (in one version of the procedure, one sample per gauge field configuration, thus $T \sim$ a few dozen) for these correlation functions. These samples are used to estimate the statistical correlation matrix between the $N$ degrees of freedom, necessary for a least-squares fit to the correlation functions.

In cosmology \cite{J}, when investigating cosmic shear, i.e., the distortion of images of distant galaxies by weak gravitational lensing caused by the large-scale structure, the basic observable is the shear two-point correlation function, computed from products of ellipticities of galaxy pairs located within given angular bins, $N \sim$ a hundred to thousand. This correlation function can be estimated from mock simulations, only $T \sim$ a few thousands, as they are computationally expensive.

In finance, one needs to know correlations between $N$ investment assets, e.g. S$\&$P500 stocks ($N \sim$ a few hundred), in order to construct optimal, well-diversified portfolios, in the spirit of Markowitz theory \cite{M,BBP,LW2017a}. We estimate these correlations from $T$ daily stock returns ($T \sim$ hundreds to thousands, due to stationarity requirements).

In genomics \cite{SS}, one analyzes expression profiles of $N$ genes ($N \sim$ thousands, from which we may select a subset of differentially expressed genes, say $N \sim$ a hundred) in order to describe similarities and functional groupings of them. Each gene is sampled $T$ times in a microarray experiment ($T \sim$ a few observation times), and the resulting data matrix $Y$, having chosen some suitable gene expression measure, is called a gene expression matrix.

In signal processing and acoustics \cite{ZSZ}, one considers an array of $N$ ($\sim$ a dozen) directional sensors, and performs so-called beamforming, combining elements of an antenna array in such a way as to have constructive/destructive interference along given directions. But (frequency-domain) beamforming algorithms, such as the Capon beamformer, require the knowledge of the correlation matrix between the array outputs, which in turn is estimated from $T$ ($\sim$ a few dozen) samples.

In anomaly detection in hyperspectral images \cite{BEA}, one has a sensor ``image'' comprised of $N$ spectral bands ($\sim$ a hundred), i.e., collected narrow ranges of the electromagnetic spectrum. When the goal is detecting anomalous pixels in such an image, one needs the Mahalanobis distance between each ($N$-dimensional) pixel and their background mean; that in turn requires the correlation matrix between the bands, estimated from a local sliding window of pixels, of length $T$ which is not too large in order to cover a homogeneous area of space ($\sim$ hundreds).

\section{Covariance matrix estimators}

\subsection{Classical regime}

This problem of estimating the covariance matrix is thus of paramount importance in very diverse fields of science, and has accordingly inspired a large body of research. Actually, a situation that is even more typical than what the above examples show, is that of $N \ll T$, meaning that one is able to collect many more samples than there are correlated entities. This is the textbook regime of \textit{classical multivariate statistics},
\begin{equation}\label{eq:classical}
    N \mbox{ finite} \quad \mbox{and} \quad T \rightarrow \infty .
\end{equation}

In this setting, the usual \textit{sample covariance estimator},
\begin{equation}\label{eq:E}
E = \frac{1}{T} Y Y^{\top} ,
\end{equation}
converges almost surely to $C$, making it a strongly consistent estimator of the population covariance matrix. In particular, the expected value of the Frobenius norm of the difference $(E - C)$, i.e., the mean squared error (MSE),
\begin{equation}\label{eq:MSE}
    L(E) = \left\langle \Vert E - C \Vert^{2} \right\rangle 
    = \left\langle \textrm{Tr} \, (E - C)^{2} \right\rangle ,
\end{equation}
tends to zero in this limit (\ref{eq:classical}). From the point of view of estimation theory, $L$ is a \textit{loss function}, and measures the quality of an estimator, here $E$. In the classical statistical regime, the loss is zero for the sample estimator $E$ (\ref{eq:E}).

\subsection{Big data regime and noise cleaning}

Yet quite often, like in the examples outlined above, one rather faces a different limit, that of both $N$ and $T$ large and of comparable magnitude,
\begin{equation}\label{eq:big_data}
    N, T \rightarrow \infty \quad \mbox{and} \quad \frac{N}{T} \rightarrow q
\end{equation}
with $q$ some positive constant. In this \textit{big data regime}, the simultaneous estimation of $\sim N^{2}$ elements of $C$ from $T \sim N$ samples leads to substantial noise (variance) in the sample estimator $E$, rendering the whole process close to meaningless. To put it in a practical context of one of the examples, if a financial manager wanted to allocate assets according to a Markowitz-type procedure, and used $E$ in the relevant formulas (which seems a very natural thing to do!), they would be optimizing their portfolio for noise! Clearly, more powerful statistical methods are required~\cite{M}.

In mathematical terms, the MSE loss (\ref{eq:MSE}) is no longer zero in the big data regime. The challenge is then to construct other estimators $\Xi$ of $C$ which have a lower value of the loss as compared to the sample estimator $E$. Such novel estimators would have some of the statistical noise \textit{cleaned}, rendering them more useful in practice. (For instance, our portfolio manager could more safely use such $\Xi$ in the Markowitz formulas.) Due to the prevalence of big data scenarios in scientific and industrial applications, and the generic nature of the estimation problem, this challenge has been met in recent years with considerable effort. We will now outline some key results, as well as add a new brick to the construction.

\subsection{Bayesian approach and linear shrinkage}

One way to search for estimators that are optimal in the regime (\ref{eq:big_data}), i.e. at finite $q$, is through an approach generally known as \textit{shrinkage}. The idea is to construct estimators with higher bias, yet much lower variance, so that the MSE (\ref{eq:MSE}) (which can be decomposed as bias$^{2}$ + variance) is in effect reduced (``shrunk'').

This concept of shrinkage has been first introduced in the problem of estimating the mean $\vec{\mu}$ of an $N$-dimensional normally-distributed variable $\vec{Y}$ with given covariance matrix $\sigma^2 \mathbb{I}$, based on a single observation $\vec{y}$. The standard least-squares (or maximum-likelihood) estimator is simply,
\begin{equation}\label{eq:mean_LS}
    \vec{\mu}_{\textrm{LS}} = \vec{y}
\end{equation}
However, \textit{James-Stein} (JS) \cite{JS} demonstrated that a biased estimator of the mean,
\begin{equation}\label{eq:mean_JS}
    \vec{\mu}_{\textrm{JS}} = \left( 1 - \frac{(N - 2) \sigma^2}{\Vert \vec{y} \Vert^2} \right) \vec{y} ,
\end{equation}
in dimension $N \geq 3$, has in fact a lower MSE than (\ref{eq:mean_LS}), precisely due to reduced variance, which more than compensates for increased bias. The existence of the JS estimator is often expressed as \textit{Stein's paradox}, that when estimating three or more parameters simultaneously, there exists a combined estimator with better MSE than any method handling them separately. The paradox's resolution lies in noticing that we are reducing the \textit{total} MSE, not that of individual components, and so the JS estimator should really be used only when this total error is of interest rather than the individual errors.

The JS estimator can be naturally understood within the \textit{empirical Bayes} approach (cf. \cite{BBP} for an extensive discussion): $\vec{\mu}$ itself is considered a random variable, with a prior distribution assumed Gaussian of zero mean and some given variance. Minimizing the MSE (\ref{eq:MSE}), where now averaging is w.r.t. the \textrm{joint} distribution of $\vec{Y}$ and $\vec{\mu}$, leads to expression (\ref{eq:mean_JS}). The choice of the prior is dictated by tractability: it is a \textit{conjugate prior} w.r.t. the Gaussian likelihood, which means that the posterior distribution is analytically calculable; thus the minimization problem of the loss $L$ can be carried out explicitly.

Let's return now to our current problem of covariance estimation. It has been precisely an analogous Bayesian reasoning that led to the first successful attempt at constructing a shrinkage estimator of $C$. Indeed, assuming the likelihood $P(Y | C)$ to be multivariate Gaussian with covariance $C$, which is a common starting point in many applications, the so-called \textit{inverse-Wishart} prior $P(C)$ turns out to be conjugate to it. Hence, loss (\ref{eq:MSE}) minimization can again be carried out analytically, and the result is the \textit{Ledoit-Wolf linear shrinkage estimator} \cite{LW2004},
\begin{equation}\label{eq:LW}
    \Xi = \alpha_{s} E + (1 - \alpha_{s}) \mathbb{I} .
\end{equation}
It interpolates (with some $\alpha_{s}$ estimated from the data) between the sample estimator $E$ and the null hypothesis, which here is the unit matrix; the null hypothesis can also be a more general \textrm{prior matrix} $C_0$, encoding a more specific prior belief, achieved by a generalized version of the inverse-Wishart prior. This simple result (\ref{eq:LW}) has been the workhorse of large-dimensional estimation, and can be found in all the applications mentioned in the introduction.

The choice of the inverse-Wishart conjugate pair is, however, dictated more by computational tractability than any insight gleaned from the data, and in fact the observed sample estimator $E$ strongly constraints the prior distribution of $C$; indeed, through the Mar\v{c}enko-Pastur law (\ref{eq:MP_density}), which we discuss later on. The recent strand of research \cite{BBP,LW2012,LW2015,LW2017b,LW2020}, initiated by \cite{LP}, thus approaches the problem of constructing shrinkage estimators in greater generality, and with marked attention to data. This we are now going to outline.

\subsection{Rotationally-invariant estimators and nonlinear shrinkage}

Consider the eigendecomposition of the sample estimator,
\begin{equation}\label{eq:E_eig}
    E = \sum_{i = 1}^{N} \lambda_{i} | \lambda_{i} \rangle \langle \lambda_{i} | .
\end{equation}
As we have repeatedly stressed, in the big data regime (\ref{eq:big_data}), the \textit{curse of dimensionality} means that both the eigenvalues and eigenvectors have high variance, thus ``concealing'' behind statistical noise the underlying nature of the true $C$. Suppose, in the first approximation, that we have no prior belief on the eigenvectors of $C$, so that our goal is merely to ``clean'' the eigenvalues. In the absence of any preferred direction (bias) in the eigenvector space, the only available basis is that of $E$. In other words, we wish to seek a MSE-optimal estimator of $C$ in the space of matrices with the eigenvectors $| \lambda_{i} \rangle$,
\begin{equation}\label{eq:RIE}
    \Xi = \sum_{i = 1}^{N} \xi_{i} | \lambda_{i} \rangle \langle \lambda_{i} | .
\end{equation}
This form is called a \textit{rotationally-invariant estimator} (RIE) \cite{BBP}, because it can alternatively be obtained from a Bayesian argument with a prior on $C$ invariant under orthogonal similarity transformations. Note that the linear shrinkage estimator (\ref{eq:LW}) also has the same sample eigenvectors, so it belongs to the RIE class; its eigenvalues are simple linear functions of the sample eigenvalues, $\xi_{i} = \alpha_{s} \lambda_{i} + 1 - \alpha_{s}$. This Ledoit-Wolf estimator, despite being better in terms of the MSE (\ref{eq:MSE}) than the plain sample estimator, is nonetheless not optimal. Indeed, we will find out that the MSE-optimal \textit{shrunk eigenvalues} $\xi_{i}$ are nonlinear functions of $\lambda_{i}$, making (\ref{eq:RIE}) an example of \textit{nonlinear shrinkage}.

\subsection{Oracle RIE}

Minimizing the MSE in this space (\ref{eq:RIE}) is a quadratic problem, and so its solution is straightforward,
\begin{equation}\label{eq:xi_oracle}
    \xi_{i} = \langle \lambda_{i} | C | \lambda_{i} \rangle = \frac{1}{N} \sum_{j = 1}^{N} c_{j} \mathcal{O} (\lambda_{i}, c_{j}) ,
\end{equation}
where we introduce the eigendecomposition of the population covariance matrix,
\begin{equation}\label{eq:C_eig}
    C = \sum_{j = 1}^{N} c_{j} | c_{j} \rangle \langle c_{j} | ,
\end{equation}
as well as the \textit{overlaps} between the eigenvectors of $E$ and $C$,
\begin{equation}\label{eq:overlaps}
    \mathcal{O}(\lambda_{i}, c_{j}) = N | \langle \lambda_{i} | c_{j} \rangle |^{2} .
\end{equation}
Note that the scalar products $\langle \lambda_{i} | c_{j} \rangle$ are of order $1 / \sqrt{N}$, and so the factor of $N$ makes the overlaps of order $1$ in the large-$N$ limit. The solution (\ref{eq:xi_oracle}) can be rewritten as an integral,
\begin{equation}\label{eq:xi_oracle_int}
    \xi_{i} = \int \textrm{d} c \rho_{C} (c) c \mathcal{O} (\lambda_{i}, c) ,
\end{equation}
making use of the \textit{density of eigenvalues} of $C$,
\begin{equation}\label{eq:C_eigval_density}
    \rho_{C} (c) = \frac{1}{N} \sum_{j = 1}^{N} \delta ( c - c_{j} ) ,
\end{equation}
where $\delta$ is the Dirac delta. This form is especially convenient in the big data limit (\ref{eq:big_data}).

The formal solution (\ref{eq:xi_oracle}) is unfortunately not useful in practice as it depends on the knowledge of $C$, which is the very object we wish to estimate! After all, the purpose of this research is uncovering from data as much information about $C$ as possible. Since it requires insight into the unknown, this solution is called the \textit{oracle estimator}.

It is however a documented phenomenon that a MSE-optimal estimator may in fact happen to be independent of the parameter being estimated. A canonical example is estimating the population variance $\sigma^{2}$ of a Gaussian random variable $Y$. If we restrict the search space to estimators of the form $c V$, where $V = \frac{1}{T} \sum_{t = 1}^{T} (Y_{t} - \bar{Y})^{2}$ is the sample variance estimator, and we are looking for a constant $c$ which minimizes the MSE, $\langle (c V - \sigma^{2})^{2} \rangle$, then we find $c = T / (T + 1)$, independently of the true $\sigma^{2}$.

And indeed, a similar miracle occurs for our oracle estimator (\ref{eq:xi_oracle}) in the big data limit (\ref{eq:big_data}). In this regime, it turns out that the right hand side can be entirely expressed as a function of the sample estimator $E$, which can of course be easily computed based on the data at hand, without any recourse to $C$. This has been first demonstrated for uncorrelated samples by Ledoit and P\'ech\'e \cite{LP}, who derived an analytic expression (\ref{eq:LP}) for the $\xi_i$'s, dependent only on $E$ rather than $C$.

The purpose of the current paper is to extend this result to samples that are correlated. Our framework will be that of \textit{random matrix theory} (RMT), as well as some ideas from the \textit{free probability} calculus. In the next section, we will review the relevant notions of these theories.

\section{Random matrix theory for covariance estimation}

\subsection{Random matrix theory transforms}

Let's start by recalling a few definitions and relations which are useful in the derivation below. For a self-adjoint random matrix, an example of which is the sample covariance estimator $E$, the fundamental object is the \textit{resolvent}, also known as the \textit{Green function}. It is a matrix-valued function of a complex argument $z$,
\begin{equation}\label{eq:G}
    G_{E}(z) = (z \mathbb{I} - E)^{-1} .
\end{equation}
To fix attention, think of $E$ as coming from the ensemble of $N \times N$ matrices of the form (\ref{eq:E}), with $Y$ a rectangular $N \times T$ random matrix. We will also assume $Y$ to have normally-distributed entries. Our problem will then turn out to be analytically feasible; but at the same time it will be argued that the results are in fact representative of a wider class of matrices; hence, this assumption is less restrictive than it appears. Now, returning to the resolvent (\ref{eq:G}) in full generality, one may suppose that the randomness of $E$ makes it a random function; however, the claim is that in the large-$N$ limit it is self-averaging, that is independent of any realization of $E$, but converging to a deterministic matrix. With a slight abuse of notation, we will keep calling that deterministic function $G_{E}(z)$. In other words, the average of the resolvent over the ensemble of $E$ can be replaced by the resolvent for a single realisation of $E$ when $N \to \infty$.

Another important object is the normalized trace of the resolvent,
\begin{equation}\label{eq:Stieltjes_finite}
    g_{E}(z) = \frac{1}{N} \textrm{Tr} \, G_{E}(z) .
\end{equation}
In the large-$N$ limit, it tends to the \textit{Stieltjes transform} of the eigenvalue density $\rho_{E}(\lambda)$,
\begin{equation}\label{eq:Stieltjes}
    g_{E}(z) = \int \textrm{d} \lambda \rho_{E} (\lambda) \frac{1}{z - \lambda} ,
\end{equation}
and thereby encodes the information about the spectrum of $E$. Indeed, evaluating $g_{E}(z)$ near the real axis, that is at $z = \lambda - i \epsilon$, with $\epsilon \to 0^+$, and applying the Sokhotski-Plemelj formula,
\begin{equation}\label{eq:SP}
\frac{1}{x - \textrm{i} 0^{+}} = \textrm{p.v.} \frac{1}{x} + \textrm{i} \pi \delta(x) ,
\end{equation}
where ``p.v.'' stands for the Cauchy principal value, yields,
\begin{equation}\label{eq:g_h_rho}
    g_{E}(\lambda - \textrm{i} 0^{+}) = \pi h_{E}(\lambda) + \textrm{i} \pi \rho_{E}(\lambda) ,
\end{equation}
where,
\begin{equation}\label{eq:h}
    h_{E} (\lambda) = \frac{1}{\pi} \textrm{p.v.} \int \textrm{d} \lambda' \rho_{E} (\lambda') \frac{1}{\lambda - \lambda'}
\end{equation}
is the \textit{Hilbert transform} of the eigenvalue density. The imaginary part of the above expression, on the other hand, is simply the eigenvalue density $\rho_{E}(\lambda)$.

A notion closely related to the resolvent is that of the \textit{moment generating function}, also called the \textit{$M$-transform}. In its matrix form, analogous to (\ref{eq:G}), it is defined as,
\begin{equation}\label{eq:M}
    M_{E}(z) = z G_{E}(z) - \mathbb{I} .
\end{equation}
Its normalized trace,
\begin{equation}\label{eq:m}
    m_{E}(z) = \frac{1}{N} \textrm{Tr} \, M_{E}(z) = z g_{E}(z) - 1
\end{equation}
has a useful property of generating, in the $1/z$-expansion, the \textit{moments} of $E$,
\begin{equation}\label{eq:mgf}
    m_{E}(z) = \sum_{k = 1}^{\infty} m_{E, k} z^{-k} ,
\end{equation}
where,
\begin{equation}\label{eq:moments}
    m_{E, k} = \left\langle \frac{1}{N} \textrm{Tr} \, E^{k} \right\rangle \stackrel{N \to \infty}{\longrightarrow} \int \textrm{d} \lambda \rho_{E} (\lambda) \lambda^{k} .
\end{equation}
In the free probability theory \cite{V,VDN}, one commonly uses instead of $m_{E}(z)$ another generating function, the \textit{$\psi$-transform},
\begin{equation}\label{eq:psi}
    \psi_{E}(z) = m_{E}(1/z) ,
\end{equation}
since it produces the moments as coefficients of the $z$-expansion, rather than $1/z$ as in (\ref{eq:mgf}).

Let's remark at this point that we have introduced our objects in pairs: a matrix version and its normalized trace version. In free probability, only the latter is relevant, and that is because there we only deal with distributions of \textit{eigenvalues}. Here we are however also concerned about \textit{eigenvectors}, and thus we need to encode the rotational information, too, which we do in the more general matrix form. This is why we distinguish between \textit{scalar} and \textit{matrix} versions of the moment-generating functions and related objects.

\subsection{Free multiplication and further transforms}
\label{ss:mult}

The normalized-trace transforms of free probability, especially the $\psi$-transform (\ref{eq:psi}), play a central role in \textit{free multiplication} \cite{V,VDN}.

We are given two large ($N \to \infty$) random matrices, $A$ and $B$; in particular we know their respective eigenvalue densities, $\rho_{A}(\lambda)$ and $\rho_{B}(\lambda)$. We assume they are \textit{free} with respect to each other; ``freeness'' is a matrix analog of statistical independence of random variables. In such a case, it turns out it is possible to derive the eigenvalue density of their matrix product, $\rho_{A B}(\lambda)$, based solely on the individual densities. To be more precise, $A$ and $B$ should be self-adjoint, so that their eigenvalues are real, and the densities are real-valued. But the product $A B$ won't then typically be self-adjoint. If $A$ is positive-definite, and $B$ has trace different from zero, then the alternative product $\sqrt{A} B \sqrt{A}$ has the same moments as $A B$, and is self-adjoint. This we will call the ``free product''.

The prescription to do so makes use of a \textit{$\chi$-transform}, which is the functional inverse of the $\psi$-transform (\ref{eq:psi}),
\begin{equation}\label{eq:chi}
    \chi_{E}(\psi_{E}(z)) = \psi_{E}(\chi_{E}(z)) = z .
\end{equation}
Furthermore, yet another useful object, the \textit{$S$-transform}, is related to the above as,
\begin{equation}\label{eq:S}
    S_{E}(z) = \frac{1 + z}{z} \chi_{E}(z) .
\end{equation}
A justification for such a seemingly arbitrary construction is that in this language the law of free multiplication becomes very straightforward, namely the $S$-transforms are simply multiplicative under the free product of matrices,
\begin{equation}\label{eq:mult}
    S_{A B}(z) = S_{A}(z) S_{B}(z) .
\end{equation}
The logic here is that the known eigenvalue densities of the constituents $A$ and $B$ are used to calculate the respective Stieltjes transforms (\ref{eq:Stieltjes}) of both matrices, then their $\psi$-transforms through the simple formula (\ref{eq:psi}), whereupon functional inversion gives the $\chi$-transforms (\ref{eq:chi}), and finally the $S$-transforms (\ref{eq:S}). After multiplying the two according to (\ref{eq:mult}), we follow the above path backward to eventually find the eigenvalue density of the free product $\sqrt{A} B \sqrt{A}$.

\subsection{The sandwich model}

Remember that our present goal is to demonstrate that the oracle estimator (\ref{eq:xi_oracle_int}), with the eigenvector overlaps (\ref{eq:overlaps}), can be expressed entirely in terms of the observable $E$, rather than the true and unknown $C$. To this end, we need to somehow \textit{model} the relationship between $C$ and $E$, or equivalently, between $C$ and the observed data $Y$. Such a model, simplistic as it may be, should be capable of capturing essential features of the data-generating process.

Recall that $Y = [y_{i t}]$ is an $N \times T$ matrix of observations, all of which we now treat as random variables. The simplest possible model is that these entries are normally-distributed, in such a way that the samples of a given entity are uncorrelated among themselves, while any two entities $i$ and $j$ are correlated according to the population matrix $C_{i j}$,
\begin{equation}\label{eq:sandwich_1}
    \langle y_{i t} y_{j s} \rangle = C_{i j} \delta_{t s} .
\end{equation}
This structure has been extensively studied. The topic of the current publication is to take it one step further in a very natural direction \cite{BJW,BJJNPZ,BJNS}: by replacing the Kronecker delta in (\ref{eq:sandwich_1}) with an arbitrary matrix $A_{t s}$, real symmetric and positive semi-definite,
\begin{equation}\label{eq:sandwich_2}
    \langle y_{i t} y_{j s} \rangle = C_{i j} A_{t s} .
\end{equation}
We assume also
\begin{equation}\label{eq:sandwich_2_norm}
\frac{1}{T} \textrm{Tr} \, A = 1 ,
\end{equation}
so that the decomposition on the right hand side is unambiguous. You may think of the dataset generated synthetically by a multivariate stochastic process (e.g., a multidimensional Ornstein-Uhlenbeck process, or a VARMA process), alternatively the dataset consisting of historic observations; $A_{t s}$ describes \textit{auto-correlations} between samples.

Note that, crucially, these auto-correlations are assumed identical for all the entities, i.e., decoupled from the cross-correlations. It would be quite interesting to extend this work to a more general coupled structure, $\langle y_{i t} y_{j s} \rangle = \mathcal{C}_{i j t s}$.

For a \textit{stationary} stochastic process, the auto-correlation matrix $A$ is a Toeplitz matrix,
\begin{equation}\label{eq:toeplitz}
    A_{t s} = a(|t - s|) ,
\end{equation}
with $a$ some function such that $a(0) = 1$ (to ensure normalized trace). A simple and often physically justified case is that of exponentially-decaying auto-correlations,
\begin{equation}\label{eq:exp_decay}
    a(t) = e^{- t / \tau} ,
\end{equation}
with auto-correlation time $\tau$. This model is a good starting point for systems exhibiting \textit{short-range} correlations between samples.

The model (\ref{eq:sandwich_2}) is simple enough to allow for an analytical solution (below), yet, as in the central limit theorem, it is believed to describe in the large-dimensional limit a broader class of distributions (not necessarily Gaussian), only perhaps without fat tails. In fact, even fat tails may be accounted for: it is known that the maximum-likelihood estimator of correlations for fat-tailed random variables is the robust \textit{Maronna estimator}; and it turns out \cite{CKP} it can be expressed in the form (\ref{eq:sandwich_2}) with a proper choice of $A$.

Yet another application \cite{TZ} of the model (\ref{eq:sandwich_2}) is that a proper choice of $A$ allows us to describe the \textit{exponentially-weighted moving average} (EWMA) estimator of $C$. Essentially, instead of treating all the samples equally as in the sample estimator $E$ (\ref{eq:E}), older samples are given lower weights than newer ones, thereby taking into account possible non-stationarity of the data. (We plan to address these cases in a separate publication.)

An attentive reader may wonder at this point how it is that while $C$ is unknown (and we are trying to estimate it), we seem to presume the knowledge of $A$. Indeed, in the latter two applications we mentioned (the Maronna and EWMA estimators), $A$ plays more of a technical role: it turns out that the structures appearing in other contexts (covariance estimation under fat tails, or samples getting obsolete due to non-stationarity, respectively), can be translated to the language of (\ref{eq:sandwich_2}) with a certain $A$. In the first example, though, we should genuinely try to estimate $A$ (e.g., via estimating $\tau$) from the data. However, an alternative way of thinking is that $\tau$ is an \textit{effective parameter}, giving our model an additional degree of freedom; that added flexibility can then be used to fit the data better. A thorough investigation of this topic is left for another paper; but already at this point we announce the logic of the matter: the resulting rotationally-invariant estimator, toward which we are working, will obviously depend on $\tau$, and in a real-world scenario that would allow us to choose this effective auto-correlation time so that the loss function (\ref{eq:MSE}) is lower than for $A = \mathbb{I}$. Interestingly, even though the MSE clearly depends on the unknown $C$, we will see a way of fitting $\tau$ even without the knowledge of $C$.

To recapitulate, we assume a simple data-generating model, a Gaussian random matrix $Y$ with the correlation structure (\ref{eq:sandwich_2}). Since a linear combination of Gaussian variables remains Gaussian, it is easy to show that an equivalent formulation is that of a linear combination of standard normal variables,
\begin{equation}\label{eq:sandwich_3}
    Y = \sqrt{C} X \sqrt{A} ,
\end{equation}
with $X = [x_{i t}]$ an $N \times T$ matrix consisting of i.i.d. $N(0, 1)$ random variables; in particular, $\langle x_{i t} x_{j s} \rangle = \delta_{i j} \delta_{t s}$. (The square roots are well-defined as $C$ and $A$ are positive semi-definite.) The modeled relation between the sample estimator $E$ and the true $C$ is therefore,
\begin{equation}\label{eq:sandwich_4}
    E = \frac{1}{T} \sqrt{C} X A X^{\top} \sqrt{C} .
\end{equation}
Its structure is that of a \textit{sandwich} consisting of several layers of matrices multiplied together.

\subsection{Mar\v{c}enko-Pastur equation}

We remind again that we are on a quest to express the oracle estimator (\ref{eq:xi_oracle}) through $E$ rather than $C$. The first step has been to set up a model between the two matrices, tractable on one hand, and realistic enough on the other. The ``sandwich model'' (\ref{eq:sandwich_4}) has been our choice, allowing not only for correlated samples (a phenomenon observed in practical applications), but also technical extensions such as the Maronna or EWMA estimators.

The second step is to express this relationship mathematically through the random matrix theory transforms defined above. We will see that this language will allow us to rewrite the eigenvector overlaps (\ref{eq:overlaps}) entirely in terms of $E$.

There are two levels to this construction. First, we have already remarked that (\ref{eq:sandwich_4}) is a product of matrices; in fact, we see that up to a certain reshuffling of factors it is a product of $C$, $A$, and the standardized \textit{Wishart random matrix},
\begin{equation}\label{eq:W}
    W = \frac{1}{T} X X^{\top} .
\end{equation}
It turns out that from the point of view of \textit{free probability} this reshuffling does not matter, as we elucidate in appendix~\ref{app:MP_scalar} (it is essentially due to the cyclic property of trace), and so the free multiplication law (\ref{eq:mult}) leads directly to \cite{BJJNPZ},
\begin{equation}\label{eq:MP_scalar_1}
    S_{E}(z) = S_{C}(z) S_{A}(q z) \frac{1}{1 + q z} .
\end{equation}
The last term is the $S$-transform of the Wishart matrix (\ref{eq:W}), $S_{W}(z) = 1/(1 + q z)$. The additional factor $q = N / T$ in the argument of $S_{A}(q z)$ comes from the difference in dimensions between $A$ and $C$.

Using elementary algebra and relations between the various transforms, (\ref{eq:MP_scalar_1}) can be rewritten in a suggestive form,
\begin{equation}\label{eq:MP_scalar_2}
    m \equiv m_{E}(z) = m_{C}(Z) , \quad \frac{Z(z)}{z} = \frac{S_{A}(q m)}{1 + q m} .
\end{equation}
In other words, the scalar (i.e., the normalized-trace version) moment-generating function (MGF) (\ref{eq:m}) of the sample estimator, evaluated at a complex argument $z$, which we denote for short $m \equiv m_{E}(z)$, is equal to the scalar MGF of the population covariance matrix, but evaluated at a different complex argument $Z$, $m_{C}(Z)$. This complex transformation $z \to Z$ depends on $m$, and does so through the $S$-transform of the auto-correlation matrix $A$. Note that we may equivalently express the right hand side through the $\chi$-transform (\ref{eq:chi}), $Z/z = \chi_{A}(q m) / q m$, which will in fact be more convenient for us. We will refer to (\ref{eq:MP_scalar_2}) as the \textit{(scalar) generalized Mar\v{c}enko-Pastur equation}.

Let's also mention that (\ref{eq:MP_scalar_2}) is sometimes spelled in the literature by explicitly writing the moment-generating function of $C$ as an integral (\ref{eq:Stieltjes}),
\begin{equation}\label{eq:MP_scalar_3}
    \int \textrm{d} \lambda \rho_{C} (\lambda) \frac{\lambda}{Z - \lambda} = m .
\end{equation}
In particular, this equation for $A = \mathbb{I}$, i.e., with $Z = z/(1 + q m)$, is the classical Mar\v{c}enko-Pastur law of 1967 \cite{MP}; that is why we call (\ref{eq:MP_scalar_2}) ``generalized''. We will however refrain from such lengthy integral expression, preferring instead the brevity offered by the various transforms we have introduced.

One way of thinking about (\ref{eq:MP_scalar_2}) is the following: Suppose one is given the true matrices $C$ and $A$; say, one generates the data from a stochastic process with these matrices as inputs. Consequently, the MGF of $C$ and the $S$-transform of $A$ can be (at least in principle) calculated. Then (\ref{eq:MP_scalar_2}) is a system of equations with unknown $m$. In some cases this system is explicitly solvable; for instance, for $C = \mathbb{I}_{N}$ and $A = \mathbb{I}_{T}$, we have $m_{C}(u) = 1/(u - 1)$ and $S_{A}(u) = 1$, directly from the definitions, and hence (\ref{eq:MP_scalar_2}) becomes a quadratic equation for $m$, $q m^2 + m (1 + q - z^2) + 1 = 0$. Solving this equation, and using the Sokhotski-Plemelj formula (\ref{eq:g_h_rho}), leads to the density of eigenvalues of $E$,
\begin{equation}\label{eq:MP_density}
    \rho_{E}(\lambda) = \frac{1}{2 \pi q \lambda} \sqrt{ (\lambda_{+} - \lambda) (\lambda - \lambda_{-}) } ,
\end{equation}
with $\lambda_{\pm} \equiv (1 \pm \sqrt{q})^{2}$. This is the famous \textit{Mar\v{c}enko-Pastur density} \cite{MP}, and it beautifully demonstrates the very problem we are trying to solve in this paper: the true correlation eigenvalue is in this case $c_{i} = 1$ (with multiplicity $N$), but the observed sample eigenvalues $\lambda_{i}$ are scattered around $1$, smeared by statistical noise, the more so the greater $q = N/T$ is, i.e., the more we are into the big data regime (and irrespective of how many samples $T$ we have collected!); the underlying eigenvalue $1$ is hidden inside this ``blob'' of sample eigenvalues, $[\lambda_{-}, \lambda_{+}]$.

Assuming the knowledge of $C$ and $A$, and calculating $m \equiv m_{E}(z)$ from them, is thus an useful perspective on the generalized Mar\v{c}enko-Pastur equation (\ref{eq:MP_scalar_2}), but this is \textit{not} the problem we are trying to solve. Rather, we know $m$, which is straightforward to compute from the observed data through the sample estimator $E$. We also suppose we know $A$, which, as we have discussed above, is either a technical construct, or can be treated as a set of effective parameters. The goal is to find $C$, which will turn the oracle estimator (\ref{eq:xi_oracle}) into an observable quantity.

An attentive reader will now notice, though, that the oracle estimator depends on the overlaps (\ref{eq:overlaps}) of the \textit{eigenvectors} of $C$ and $E$, which the \textit{scalar} Mar\v{c}enko-Pastur equation is not capable of capturing. Rather, we need an equation relating the \textit{matrix} versions of the transforms in question. Interestingly enough, such an equation exists, and is in fact completely parallel to its scalar version (\ref{eq:MP_scalar_2}),
\begin{equation}\label{eq:MP_matrix}
    M_{E}(z) = M_{C}(Z) ,
\end{equation}
with complex numbers $Z$ and $z$ related in the exact same way as before. This is an $N \times N$ matrix equation, expressed through the matrix moment-generating functions (\ref{eq:M}). Of course, the previous scalar version follows by taking normalized trace of both sides. We will call (\ref{eq:MP_matrix}) the \textit{matrix generalized Mar\v{c}enko-Pastur equation}. It has first been derived in \cite{BJW} using diagrammatic methods; \cite{BBP} presents another derivation via the replica trick and the low-rank orthogonal Harish-Chandra-Itzykson-Zuber integral.

\section{Nonlinear shrinkage for correlated samples}

\subsection{Shrinkage estimator for the sandwich model}

We are now ready to return to the problem at hand, which is to express the oracle estimator entirely through $E$. The key point is to rewrite the eigenvector overlaps (\ref{eq:overlaps}) in the language of the above transforms, and to apply the generalized Mar\v{c}enko-Pastur equation (\ref{eq:MP_matrix}), (\ref{eq:MP_scalar_2}) to translate all the references to transforms of $C$ to those of $E$ only.

To this end, write the matrix resolvent (\ref{eq:G}) of $E$ through the sample eigenvalues and eigenvectors (\ref{eq:E_eig}), in an integral form weighted by the eigenvalue density,
\begin{equation}\label{eq:calc_1}
    G_{E}(z) = N \int \textrm{d} \lambda' \rho_{E}(\lambda') \frac{1}{z - \lambda'} | \lambda' \rangle \langle \lambda' | .
\end{equation}
Evaluate this expression near the real axis, at $z = \lambda - \textrm{i} 0^{+}$, and retain the imaginary part of the result; the Sokhotski–Plemelj formula (\ref{eq:SP}) implies,
\begin{equation}\label{eq:calc_2}
    \frac{1}{N \pi \rho_{E}(\lambda)} \textrm{Im} \, G_{E}(\lambda - \textrm{i} 0^{+}) = | \lambda \rangle \langle \lambda | .
\end{equation}
Insert this equality inside a scalar product with an eigenvector of $C$,
\begin{equation}\label{eq:calc_3}
    \frac{1}{\pi \rho_{E}(\lambda)} \textrm{Im} \, \langle c | G_{E}(\lambda - \textrm{i} 0^{+}) | c \rangle = \mathcal{O}(\lambda, c) .
\end{equation}
We have thereby expressed the eigenvector overlap through the matrix resolvent of $E$; still, however, there is an explicit reference to an eigenvector of $C$, which we will now remove in favor of some transform of $C$.

Indeed, the matrix resolvent of $E$ evaluated at any complex $z$ is equivalent, thanks to the matrix Mar\v{c}enko-Pastur equation (\ref{eq:MP_matrix}), to the matrix resolvent of $C$ at the transformed $Z = Z(z)$ (\ref{eq:MP_scalar_2}), which has in turn a simple behavior when surrounding it with an eigenvector of $C$,
\begin{equation}\label{eq:calc_4}
    \langle c | G_{E}(z) | c \rangle = \frac{Z}{z} \langle c | G_{C}(Z) | c \rangle = \frac{Z}{z} \frac{1}{Z - c} .
\end{equation}
Inserting this to (\ref{eq:calc_3}),
\begin{equation}\label{eq:calc_5}
    \frac{1}{\pi \rho_{E}(\lambda)} \textrm{Im} \, \left. \frac{Z(z)}{z} \frac{1}{Z(z) - c} \right|_{z = \lambda - \textrm{i} 0^{+}} = \mathcal{O}(\lambda, c) .
\end{equation}

In the oracle estimator's generic eigenvalue $\xi$ (\ref{eq:xi_oracle_int}), the overlap appears multiplied by the corresponding eigenvalue $c$, and integrated over $c$ with the proper eigenvalue density, $\rho_{C}(c)$. This operation, applied to the left hand side of (\ref{eq:calc_5}), produces according to (\ref{eq:m}), (\ref{eq:Stieltjes}) the scalar moment-generating function of $C$ evaluated at the complex argument $Z$,
\begin{equation}\label{eq:calc_6}
    \frac{1}{\pi \rho_{E}(\lambda)} \textrm{Im} \, \left. \frac{Z(z)}{z} m_{C}(Z(z)) \right|_{z = \lambda - \textrm{i} 0^{+}} = \xi .
\end{equation}

The final step is to replace in (\ref{eq:calc_6}) the scalar MGF of $C$ at argument $Z$ by the scalar MGF of $E$ at argument $z$, according to the scalar Mar\v{c}enko-Pastur equation (\ref{eq:MP_scalar_2}),
\begin{equation}\label{eq:calc_7}
    \frac{1}{\pi \rho_{E}(\lambda)} \textrm{Im} \, \left. \frac{Z(z)}{z} m_{E}(z) \right|_{z = \lambda - \textrm{i} 0^{+}} = \xi .
\end{equation}
And this is precisely what we set out to achieve! The oracle eigenvalue $\xi$, previously accessible only through the unknown $C$, is now expressed solely in terms of the observable $E$, through the scalar MGF, $m = m_{E}(\lambda - \textrm{i} 0^{+})$, present in (\ref{eq:calc_7}) either explicitly, or inside $Z / z = S_{A}(q m) / (1 + q m) = \chi_{A}(q m) / q m$ (\ref{eq:MP_scalar_2}).

Indeed, let's rewrite (\ref{eq:calc_7}) even more appealingly. For any observed sample eigenvalue $\lambda_{i}$, for $i = 1, \ldots, N$, choose as a basic quantity,
\begin{equation}\label{eq:main_1}
    u_{i} \equiv q m_{E}(\lambda_{i} - \textrm{i} 0^{+}) = \alpha_{i} + \textrm{i} \beta_{i} ,
\end{equation}
where since we are close to the real axis, the Sokhotski–Plemelj formula implies (\ref{eq:m}), (\ref{eq:g_h_rho}),
\begin{equation}\label{eq:main_2}
\alpha_{i} = q (\pi \lambda_{i} h_{E}(\lambda_{i}) - 1) , \quad \beta_{i} = q \pi \lambda_{i} \rho_{E}(\lambda_{i}) ,
\end{equation}
through the directly observable Hilbert transform and density of eigenvalues of the sample estimator $E$.

In this notation, (\ref{eq:calc_7}) becomes the main formula of this paper, the \textit{nonlinear shrinkage for correlated samples},
\begin{equation}\label{eq:main_3}
    \frac{\xi_{i}}{\lambda_{i}} = \frac{\textrm{Im} \, \chi_{A}(u_{i})}{\textrm{Im} \, u_{i}} ,
\end{equation}
where recall the $\chi$-transform (\ref{eq:chi}) of $A$. We note also that since the $S$-transform (\ref{eq:S}) is probably better known than the $\chi$-transform, we may replace the latter in (\ref{eq:main_3}) by $u_{i} S_{A}(u_{i}) / (1 + u_{i})$.

We stress again the logic behind this formula:

\begin{itemize}

\item We perform $T$ observations of some $N$ given entities, and collect these measurements in an $N \times T$ data matrix $Y$.

\item From this dataset, we calculate the standard sample estimator $E$ (\ref{eq:E}), and diagonalize it, obtaining in particular a set of sample eigenvalues $\lambda_{i}$, $i = 1, \ldots, N$. As we have extensively discussed, in the big data regime (\ref{eq:big_data}), these eigenvalues have a crucial component of statistical noise, making them unreliable for any estimation purposes.

\item From this set of sample eigenvalues, we first estimate the Hilbert transform $h_{E} (\lambda)$ and density of eigenvalues $\rho_{E} (\lambda)$. We will discuss a relevant procedure below, based on the kernel method.

\item For any given sample eigenvalue $\lambda_{i}$, we thus readily calculate the corresponding $\alpha_{i}$ and $\beta_{i}$ (\ref{eq:main_2}), and so $u_{i} = \alpha_{i} + \textrm{i} \beta_{i}$ (\ref{eq:main_1}).

\item Having assumed some model of the auto-correlations $A$ (see below), we know (at least in principle) the $\chi$- (equivalently, $S$-) transform of $A$. Thus (\ref{eq:main_3}) gives us explicitly the \textit{cleaned} (a.k.a. \textit{shrunk}) eigenvalue $\xi_{i}$. This is an eigenvalue of an optimal, w.r.t. the MSE loss function (\ref{eq:MSE}), rotationally-invariant estimator (RIE) (\ref{eq:RIE}). Since the left hand side of (\ref{eq:main_3}) is the ratio $\xi_{i} / \lambda_{i}$, the right hand side may be called a \textit{shrinkage factor}.

\end{itemize}

One sanity check is that for $q \to 0$ we are moving from the big data regime (\ref{eq:big_data}) to the classical regime (\ref{eq:classical}), in which case we should have $\Xi = E$, as the sample estimator is then optimal (and of course $E$ belongs to the RIE class). Indeed, in this limit, $u_{i} \to 0$. An expansion of the $S$-transform around zero is known \cite{BSR}, and when coupled with the assumed normalization condition of $A$ (\ref{eq:sandwich_2_norm}), it implies $S_{A}(u_{i}) \to 1$, hence $\chi_{A}(u_{i}) \to u_{i}$, and so $\xi_{i} \to \lambda_{i}$, as expected.

\subsection{The Ledoit-P\'{e}ch\'{e} formula}

Historically the first nonlinear shrinkage formula, and a cornerstone of the theory we are elaborating here, has been the \textit{Ledoit-P\'{e}ch\'{e} (LP) shrinkage} \cite{LP} (cf. also \cite{BBP}), derived for the sandwich model in the case of no correlations between samples (\ref{eq:sandwich_1}).

To verify that our formula (\ref{eq:main_3}) reduces appropriately, set $A = \mathbb{I}$. Directly from the definitions (\ref{eq:G}), (\ref{eq:M}), (\ref{eq:m}), (\ref{eq:psi}), the scalar $\psi$-transform of $A$ is $\psi_{A}(z) = z/(1 - z)$, from which the $\chi$-transform (\ref{eq:chi}) follows by functional inversion, $\chi_{A}(z) = z/(z + 1)$; note that the $S$-transform (\ref{eq:S}) is simply $S_{A}(z) = 1$. The shrinkage factor (\ref{eq:main_3}) thus becomes $\textrm{Im} \, (u_{i} / (1 + u_{i})) / \textrm{Im} \, u_{i}$, that is,
\begin{equation}\label{eq:LP}
    \frac{\xi_{i}}{\lambda_{i}} = \frac{1}{| 1 + u_{i} |^{2}} = \frac{1}{(\alpha_{i} + 1)^{2} + \beta_{i}^{2}} ,
\end{equation}
which is indeed the celebrated Ledoit-P\'{e}ch\'{e} formula.

\subsection{Exponentially-decaying auto-correlations}

A non-trivial example of a model of correlations between the samples, often relevant for situations of short-ranged dependence between the steps of a sampling process, is that of an exponentially-decaying function (\ref{eq:exp_decay}), with auto-correlation time $\tau$. It is a straightforward exercise \cite{BJW,BJJNPZ} to derive the $\chi$-transform of $A$,
\begin{equation}\label{eq:chi_exp_decay}
    \chi_{A}(z) = \frac{1}{\gamma + \sqrt{\gamma^{2} - 1 + 1 / z^{2}}} ,
\end{equation}
with $\gamma = \coth(1 / \tau)$. The shrinkage factor (\ref{eq:main_3}) for this case follows immediately.

There are other interesting models of auto-correlations, such as those generated by general $\textrm{VARMA}(r_{1}, r_{2})$ stochastic processes,
\begin{equation}\label{eq:VARMA}
    Y_{i, a} = \sum_{\beta = 1}^{r_{1}} b_{\beta} Y_{i, a - \beta} + \sum_{\alpha = 0}^{r_{2}} a_{\alpha} \epsilon_{i, a - \alpha} ,
\end{equation}
with $\epsilon$ a matrix of i.i.d. standard Gaussian (or some other distribution, say Student-t) variables. Such a process has $(r_{2} + 1)$ ``MA'' (moving average) parameters $a_{\alpha}$, and $r_{1}$ ``AR'' (auto-regressive) parameters $b_{\beta}$. Note that the exponential decay is a case of VAR(1) with a particular set of parameters, $b_{1} = \textrm{e}^{-1 / \tau}$ and $a_{0} = \sqrt{1 - b_{1}^{2}}$.

Moreover, as we have alluded to above, specific models of $A$ appear for the Maronna or EWMA estimators. These issues are left for another publication.

\subsection{Numerical considerations: kernel density estimation}

As mentioned before, our shrinkage formula (\ref{eq:main_3}) for a general sandwich model (\ref{eq:sandwich_2}) is complete except for one important component, that is a numerical method of estimating from the observed dataset the Hilbert transform and density of eigenvalues of $E$, which are inputs to the basic variable $u_{i}$ (\ref{eq:main_1}), (\ref{eq:main_2}).

The problem is that the eigenvalue density is approximated by a numerical histogram, and one cannot use it directly as an input for the Hilbert transform, which is an integral transform (\ref{eq:h}). Several ideas have been proposed to alleviate this difficulty.

A very simple procedure is to choose a small but finite $\epsilon$ in (\ref{eq:main_1}), and approximate $u_{i} \approx q m_{E}(\lambda_{i} - \textrm{i} \epsilon)$. In particular, \cite{BBP} uses this approach with $\epsilon = N^{-1/2}$, and reports satisfactory behavior. In our numerical experiments, we have however found this algorithm very sensitive to the choice of $\epsilon$, and generally unstable.

For a time, a standard and comprehensive numerical solution (for uncorrelated samples) consisted of quite a complicated scheme termed \textit{inverse QuEST} (= ``Quantized Eigenvalues Sampling Transform'') by Ledoit and Wolf \cite{LW2017b}. In essence, it encompasses the following steps: First, suppose we know the true covariance eigenvalues $c_{1}, \ldots, c_{N}$. The scalar Mar\v{c}enko-Pastur relation (\ref{eq:MP_scalar_2}), for $z = \lambda - \textrm{i} 0^{+}$, can be manipulated to yield the sample eigenvalue density $\rho_{E}(\lambda)$ in a parametric form, dependent on the $c_{i}$'s. (This is the same train of thought that led to the classical Mar\v{c}enko-Pastur density formula (\ref{eq:MP_density}).) Second, we integrate the density numerically to find the cumulative distribution function, $\textrm{CDF}(\lambda) = \int_{0}^{\lambda} \textrm{d} \lambda' \rho_{E}(\lambda')$. Third, we invert it numerically to find the quantile function, i.e., from $\textrm{CDF}(\lambda) = p$ to $\lambda = Q(p)$. Fourth, an assumption is made that the observed eigenvalues are distributed according to this quantile function, i.e., that we can approximate their positions by $\tilde{\lambda}_{i} = Q(i / N)$. The above algorithm (already quite involved) thus estimates the sample eigenvalues based on known population eigenvalues, $\tilde{\lambda}_{i}(c_{1}, \ldots, c_{N})$; this is the \textit{QuEST function}.

Now the ``inverse'' part of the prescription is the following: In reality, we do not know the $c_{i}$'s, rather we observe the $\lambda_{i}$'s. We estimate the former by minimizing the mean squared error between the QuEST estimates and the measured values,
\begin{equation}\label{eq:inverse_quest}
    (\tilde{c}_{1}, \ldots, \tilde{c}_{N}) = \textrm{argmin} \frac{1}{N} \sum_{i = 1}^{N} \left( \tilde{\lambda}_{i}(c_{1}, \ldots, c_{N}) - \lambda_{i} \right)^{2} .
\end{equation}
Finally, we calculate $\tilde{\lambda}_{i}(\tilde{c}_{1}, \ldots, \tilde{c}_{N})$, from which $u_{i} = q m_{E}(\tilde{\lambda}_{i})$, without a regularizer. As we see, the Ledoit-Wolf inverse QuEST is a complicated algorithm (and we have not mentioned additional numerical intricacies that need to be addressed), but provides a very solid and stable solution.

However, recently Ledoit and Wolf \cite{LW2020} proposed a \textit{kernel method} that is computationally straightforward, stable, and exceptionally easy to use in practice. Both the Hilbert transform and the eigenvalue density are approximated as a sum of $N$ \textit{kernels} set up around each $\lambda_{i}$. That is, for the density we write,
\begin{equation}\label{eq:kernel_rho}
    \rho_{E}(\lambda) \approx \frac{1}{N} \sum_{i = 1}^{N} \frac{1}{\mathfrak{b}_{i} \lambda_{i}} \mathfrak{r} \left( \frac{\lambda - \lambda_{i}}{\mathfrak{b}_{i} \lambda_{i}} \right) .
\end{equation}
Here $\mathfrak{r}$ is a kernel function, assumed real, non-negative, normalized ($\int \textrm{d} x \, \mathfrak{r}(x) = 1$), centered ($\int \textrm{d} x \, x \, \mathfrak{r}(x) = 0$), and with unit width ($\int \textrm{d} x \, x^{2} \mathfrak{r}(x) = 1$). The parameters $\mathfrak{b}_{i}$ (\textit{bandwidths}) are local scale factors, determining the width of the individual peaks. They should be chosen to make the neighbouring peaks overlap. Ledoit and Wolf suggest to select them identical for all the eigenvalues (i.e., independent of $i$), and equal to
\begin{equation}\label{eq:kernel_bandwidth}
    \mathfrak{b} = T^{- 1 / 3} .
\end{equation}
It is a purely heuristic choice, but we too have found it stable and accurate.

The Hilbert transform (\ref{eq:h}) is a linear operation on the eigenvalue density, hence it transforms the sum (\ref{eq:kernel_rho}) into another sum,
\begin{equation}\label{eq:kernel_h}
    h_{E}(\lambda) \approx \frac{1}{N} \sum_{i = 1}^{N} \frac{1}{\mathfrak{b}_{i} \lambda_{i}} \mathfrak{h} \left( \frac{\lambda - \lambda_{i}}{\mathfrak{b}_{i} \lambda_{i}} \right) .
\end{equation}
where $\mathfrak{h}$ is the Hilbert transform of $\mathfrak{r}$.

The idea is to choose the latter such that its Hilbert transform is analytically tractable and easy to implement. A good candidate is the \textit{Epanechnikov kernel}, essentially the non-negative part (denoted $x^{+} \equiv \max(x, 0)$) of a parabola,
\begin{equation}\label{eq:kernel_epanechnikov_rho}
    \mathfrak{r}(x) = \frac{3}{4 \sqrt{5}} \left( 1 - \frac{x^{2}}{5} \right)^{+} .
\end{equation}
Its Hilbert transform (\ref{eq:h}) reads,
\begin{equation}\label{eq:kernel_epanechnikov_h}
    \mathfrak{h}(x) = \frac{3 x}{10 \pi} - \frac{3}{4 \sqrt{5} \pi} \left( 1 - \frac{x^{2}}{5} \right) \log \left| \frac{x - \sqrt{5}}{x + \sqrt{5}} \right| ,
\end{equation}
where the second term is understood to be zero for $x = \pm \sqrt{5}$.

To sum up, we have straightforward estimates of the sample eigenvalue density (\ref{eq:kernel_rho}) and Hilbert transform (\ref{eq:kernel_h}), with the heuristic bandwidth (\ref{eq:kernel_bandwidth}), and the Epanechnikov kernel (\ref{eq:kernel_epanechnikov_rho}), (\ref{eq:kernel_epanechnikov_h}). They are inserted directly into the inputs (\ref{eq:main_2}) to the main shrinkage formula (\ref{eq:main_3}).

\section{Oracle estimation by cross-validation}

\subsection{Moving-window cross-validation}

We have now completed the main derivation of the paper, with an essentially \textit{analytical} formula for shrunk eigenvalues; it does require numerical components in the form of the Epanechnikov kernel estimation, as well as some further fitting procedure described below, needed in realistic situations, but the expression (\ref{eq:main_3}) itself is explicit. There exists, however, a purely \textit{numerical} alternative construction that allows one to estimate the oracle eigenvalues directly from the dataset $Y$, with no complicated random matrix theory in the process, and with excellent performance! We will now outline the method, and discuss its applicability. It has been introduced in sec.~8.2. of~\cite{BBP}, and is a certain modification of the method by Bartz~\cite{B}.

Recall that we have started our journey from the oracle estimator (\ref{eq:xi_oracle}), minimizing the Frobenius norm between the true $C$ and the sought-for $\Xi$ in the space of rotationally invariant estimators (\ref{eq:RIE}). The solution is simple (as the problem is quadratic), $\xi_{i} = \langle \lambda_{i} | C | \lambda_{i} \rangle$, but depends on $C$, which is in principle unknown. The machinery of random matrix theory in the big data limit (\ref{eq:big_data}) is then invoked to show that the actual dependence is solely on observable quantities, the sample eigenvalues $\lambda_{i}$, via one of the shrinkage formulas discussed in this paper.

This oracle formula can however be estimated by a numerical procedure of a \textit{moving-window cross-validation}. First, imagine we have collected some more samples, and now have $T_{\textrm{total}} > T$ of them. Consider now a series of $K$ pairs of consecutive moving windows. Each such pair consists of a ``training'' (``in-sample'') window of length $T$, and a ``testing'' (``out-of-sample'') window of some length $T_{\textrm{out}}$. This pair of windows of length $(T + T_{\textrm{out}})$ we keep shifting $K$ times through the whole dataset; this in particular implies, $K = \lfloor (T_{\textrm{total}} - T) / T_{\textrm{out}} \rfloor$. For the $\mu$-th fold ($\mu = 0, \ldots, K - 1$) denote $t_{\mu} = T + \mu T_{\textrm{out}} + 1$; following the convention used in the paper that the temporal index starts from $1$, we see that the $\mu$-th test fold has indices from $t_{\mu}$ up to $(t_{\mu} + T_{\textrm{out}} - 1)$ (both inclusive), while the $\mu$-th train fold stretches between $(t_{\mu} - T)$ and $(t_{\mu} - 1)$.

For each fold $\mu$, we denote by $E^{\textrm{train}, \mu}$ the sample estimator calculated on the current train fold, and in particular $| \lambda^{\textrm{train}, \mu}_{i} \rangle$ its eigenvectors. On the other hand, by $E^{\textrm{test}, \mu}$ denote the sample estimator calculated on the current test fold.

The claim then is that the oracle eigenvalues can be estimated by the mean over all the folds of the average test sample estimator in the state given by any train sample eigenvector,
\begin{equation}\label{eq:xi_oracle_mwcv}
    \xi^{\textrm{mwcv}}_{i} = \frac{1}{K} \sum_{\mu = 0}^{K - 1} \langle \lambda^{\textrm{train}, \mu}_{i} | E^{\textrm{test}, \mu} | \lambda^{\textrm{train}, \mu}_{i} \rangle .
\end{equation}
In other words, the numerical prescription consists of repeating $K$ times a calculation of the (current) train sample estimator, its diagonalization, then calculation of the (current) test sample estimator, and finally taking the relevant scalar product.

An intuition behind this formula is that the unknown $C$ in the oracle formula can be approximated by the \textit{out-of-sample}, i.e., \textit{unknown} from the point of view of an in-sample observer, estimator.

As a side note, let us make the following remark: The test sample estimator is derived from a dataset of shape $N \times T_{\textrm{out}}$, i.e., with a different noise ratio than $q = N/T$. One might wonder if this mismatch leads then to correct estimation. An intuition why this should not matter is that the test sample estimator corresponds in the oracle formula to the true $C$, which after all knows nothing about $q$.

Let us also stress the importance of performing cross-validation that preserves the time ordering, i.e., by moving windows. This is crucial especially when we believe auto-correlations are present in the system. The method of Bartz, mentioned in the beginning, is exactly identical except that the train and test folds are chosen randomly from among the samples, thus leveling out any temporal dependence.

\subsection{Isotonic regression}

The moving-window cross-validation-estimated oracle eigenvalues (\ref{eq:xi_oracle_mwcv}) turn out to have a significant variance; this will be visually clear in the figures below, where they form a broad cloud of scatter points.

The idea of~\cite{TZ} is extraordinarily simple and effective: fit \textit{isotonic regression} to the estimated oracle eigenvalues, i.e., a monotonic function closest to the observations. We shall see that such a simple estimator is very hard to beat! Nonetheless, we shall also demonstrate that our VARMA shrinkage may be superior, once a proper search for its parameters is executed.

\section{Examples}

All the functionalities we have expounded on in this article, including the Epanechnikov kernel estimation of $h_{E}(\lambda)$ and $\rho_{E}(\lambda)$, and computing the shrunk/cleaned eigenvalues (\ref{eq:main_3}), under several sensible models of $A$, as well as the cross-validation and isotonic regression procedures for the oracle eigenvalues, have been collected in an open-source Python module called \texttt{shrinkage}, available at \texttt{https://github.com/yedrek/shrinkage}. It is versatile, extensible, and, we hope, easy to use~\cite{shrinkage}.

Below we present a few examples produced by this library, at version \texttt{1.1.0}. We work here with synthetic data, that is, we randomly generate the data matrix $Y$ of dimensions $N = 500$, $T = 1000$ according to some given models of $C$ and $A$.

Since the true $C$ is known, we will be able to quantify the efficiency of various shrinkage methods. We will measure it by the mean squared error (\ref{eq:MSE}); more precisely, for a given shrinkage estimator $\Xi$ we will calculate its \textit{Frobenius ratio},
\begin{equation}\label{eq:Frobenius}
    \mathcal{F} = \frac{\textrm{Tr} \, (\Xi - C)^{2}}{\textrm{Tr} \, (E - C)^{2}} .
\end{equation}
The lower it is, the better the cleaning scheme; in other words, $\Xi$ should be ``closer'' to $C$ than $E$ is. The Frobenius ratios found in the analysis to follow are summarized in table~\ref{tab:Frobenius}.

\begin{table}[h!]
\begin{center}
\begin{tabular}{ |c|c|c|c| }
\hline
 & Example 1 & Example 2 & Example 3 \\
\hline
oracle mwcv & 13\% & 23\% & 34\% \\
oracle mwcv isotonic & 12\% & 22\% & 34\% \\
LP & 53\% & 33\% & 60\% \\
effective LP & 33\% & 26\% & 37\% \\
VARMA fit & 12\% & 24\% & 27\% (*) \\
\hline
\end{tabular}
\end{center}
\caption{Frobenius ratios for the three examples of synthetic data considered, and various shrinkage methods. The (*) means that while the fit of VARMA shrinkage parameters in Examples 1 and 2 is that of an exp-decay, in Example 3 we fit a three-parameter VARMA(1, 1) model.}
\label{tab:Frobenius}
\end{table}

\begin{figure*}
    \centering
    \begin{subfigure}[b]{0.475\textwidth}
        \centering
        \includegraphics[width=\textwidth]{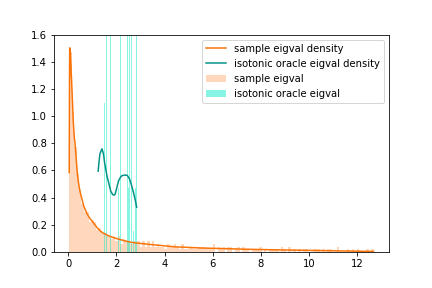}
        \caption[]{Shrinkage model: Isotonic regression fitted to cross-validation-estimated oracle eigenvalues, with Frobenius ratio 12\%.}
        \label{fig:example_1_hist_lambdas_xi_oracle_mwcv_iso}
    \end{subfigure}
    \hfill
    \begin{subfigure}[b]{0.475\textwidth}  
        \centering 
        \includegraphics[width=\textwidth]{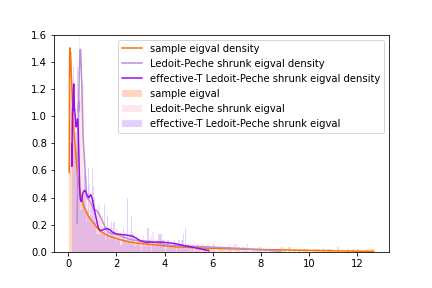}
        \caption[]{Shrinkage model: Ledoit-P\'{e}ch\'{e}, with Frobenius ratio 53\%. Ledoit-P\'{e}ch\'{e} with ``effective'' $T_{\textrm{eff}} = 426$, and Frobenius ratio 33\%.}
        \label{fig:example_1_hist_lambdas_xi_LP_xi_LP_eff}
    \end{subfigure}
    \vskip\baselineskip
    \begin{subfigure}[b]{0.475\textwidth}   
        \centering 
        \includegraphics[width=\textwidth]{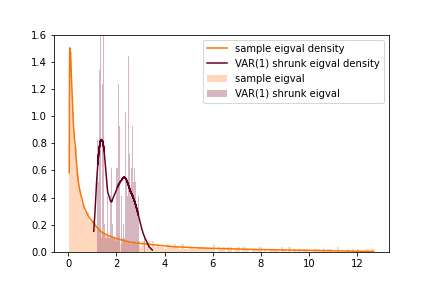}
        \caption[]{Shrinkage model: exp-decay with fitted $\tau = 2.9$, with Frobenius ratio 12\%.}
        \label{fig:example_1_hist_lambdas_xi_VAR_1_fit}
    \end{subfigure}
    \hfill
    \begin{subfigure}[b]{0.475\textwidth}   
        \centering 
        \includegraphics[width=\textwidth]{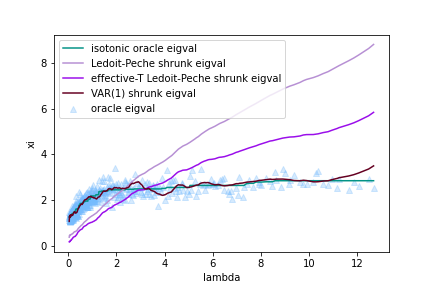}
        \caption[]{$\xi(\lambda)$}
        \label{fig:example_1_plot}
    \end{subfigure}
    \caption[]{Synthetic data model: Gaussian distribution, $C$ with eigenvalues $1$ and $3$ in 50\% proportions, $A$ exp-decay with $\tau = 3$. Figures \ref{fig:example_1_hist_lambdas_xi_oracle_mwcv_iso}, \ref{fig:example_1_hist_lambdas_xi_LP_xi_LP_eff}, \ref{fig:example_1_hist_lambdas_xi_VAR_1_fit} are normalized histograms of the sample and shrunk eigenvalues (bars), and the corresponding eigenvalue densities (lines). Figure \ref{fig:example_1_plot} contains plots of the shrinkage relationship $\xi(\lambda)$ considered in the previous figures, as well the cross-validation-estimated oracle eigenvalues (triangles).}
    \label{fig:example_1}
\end{figure*}

\begin{figure*}
    \centering
    \begin{subfigure}[b]{0.475\textwidth}
        \centering
        \includegraphics[width=\textwidth]{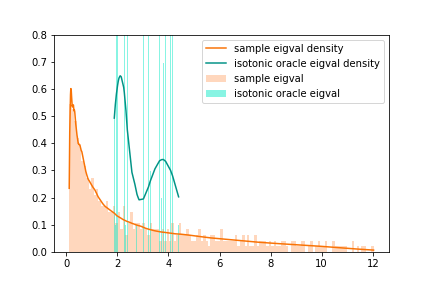}
        \caption[]{Shrinkage model: Isotonic regression fitted to cross-validation-estimated oracle eigenvalues, with Frobenius ratio 22\%.}
        \label{fig:example_2_hist_lambdas_xi_oracle_mwcv_iso}
    \end{subfigure}
    \hfill
    \begin{subfigure}[b]{0.475\textwidth}  
        \centering 
        \includegraphics[width=\textwidth]{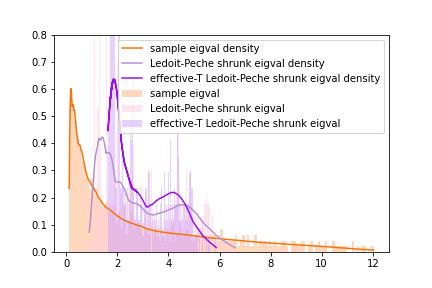}
        \caption[]{Shrinkage model: Ledoit-P\'{e}ch\'{e}, with Frobenius ratio 33\%. Ledoit-P\'{e}ch\'{e} with ``effective'' $T_{\textrm{eff}} = 811$, and Frobenius ratio 26\%.}
        \label{fig:example_2_hist_lambdas_xi_LP_xi_LP_eff}
    \end{subfigure}
    \vskip\baselineskip
    \begin{subfigure}[b]{0.475\textwidth}   
        \centering 
        \includegraphics[width=\textwidth]{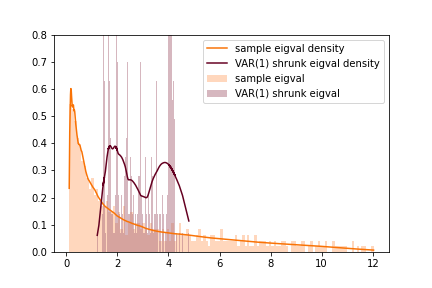}
        \caption[]{Shrinkage model: exp-decay with fitted $\tau = 1.1$, with Frobenius ratio 24\%.}
        \label{fig:example_2_hist_lambdas_xi_VAR_1_fit}
    \end{subfigure}
    \hfill
    \begin{subfigure}[b]{0.475\textwidth}   
        \centering 
        \includegraphics[width=\textwidth]{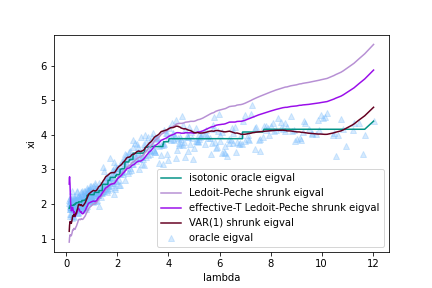}
        \caption[]{$\xi(\lambda)$}
        \label{fig:example_2_plot}
    \end{subfigure}
    \caption[]{Synthetic data model: Student-t distribution with $5$ degrees of freedom, $C$ with eigenvalues $1$ and $3$ in 50\% proportions, $A$ that of a VMA(1) model with $a_{0} = 0.8$, $a_{1} = 0.5$. Figures \ref{fig:example_2_hist_lambdas_xi_oracle_mwcv_iso}, \ref{fig:example_2_hist_lambdas_xi_LP_xi_LP_eff}, \ref{fig:example_2_hist_lambdas_xi_VAR_1_fit} are normalized histograms of the sample and shrunk eigenvalues (bars), and the corresponding eigenvalue densities (lines). Figure \ref{fig:example_2_plot} contains plots of the shrinkage relationship $\xi(\lambda)$ considered in the previous figures, as well the cross-validation-estimated oracle eigenvalues (triangles).}
    \label{fig:example_2}
\end{figure*}

\begin{figure*}
    \centering
    \begin{subfigure}[b]{0.475\textwidth}
        \centering
        \includegraphics[width=\textwidth]{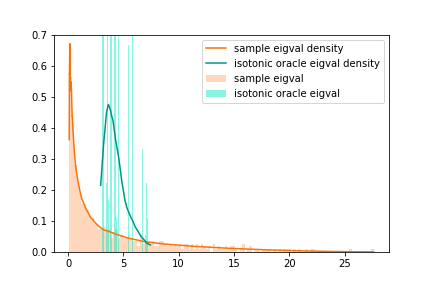}
        \caption[]{Shrinkage model: Isotonic regression fitted to cross-validation-estimated oracle eigenvalues, with Frobenius ratio 34\%.}
        \label{fig:example_3_hist_lambdas_xi_oracle_mwcv_iso}
    \end{subfigure}
    \hfill
    \begin{subfigure}[b]{0.475\textwidth}  
        \centering 
        \includegraphics[width=\textwidth]{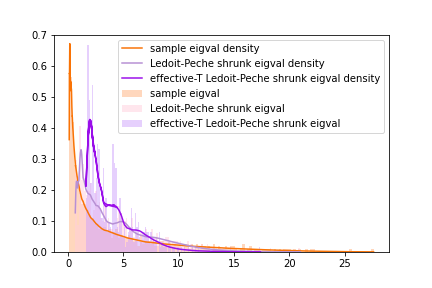}
        \caption[]{Shrinkage model: Ledoit-P\'{e}ch\'{e}, with Frobenius ratio 60\%. Ledoit-P\'{e}ch\'{e} with ``effective'' $T_{\textrm{eff}} = 565$, and Frobenius ratio 37\%.}
        \label{fig:example_3_hist_lambdas_xi_LP_xi_LP_eff}
    \end{subfigure}
    \vskip\baselineskip
    \begin{subfigure}[b]{0.475\textwidth}   
        \centering 
        \includegraphics[width=\textwidth]{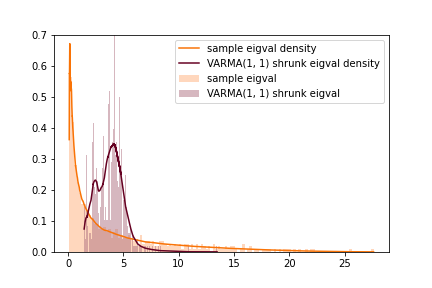}
        \caption[]{Shrinkage model: VARMA(1, 1) with fitted $a_{0} = 0.8$, $a_{1} = 0.2$, $b_{1} = 0.5$, with Frobenius ratio 27\%.}
        \label{fig:example_3_hist_lambdas_xi_VARMA_1_1_fit}
    \end{subfigure}
    \hfill
    \begin{subfigure}[b]{0.475\textwidth}   
        \centering 
        \includegraphics[width=\textwidth]{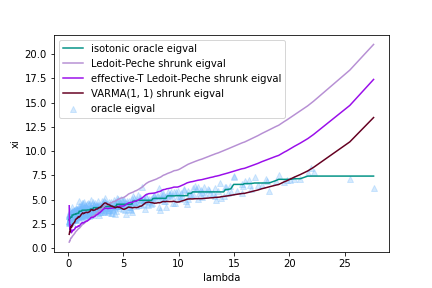}
        \caption[]{$\xi(\lambda)$}
        \label{fig:example_3_plot}
    \end{subfigure}
    \caption[]{Synthetic data model: Student-t distribution with $3$ degrees of freedom, $C$ with eigenvalues following the inverse-Wishart distribution with $\kappa = 2$, $A$ that of a VARMA(1, 1) model with $a_{0} = 0.8$, $a_{1} = 0.5$, $b_{1} = 0.4$. Figures \ref{fig:example_3_hist_lambdas_xi_oracle_mwcv_iso}, \ref{fig:example_3_hist_lambdas_xi_LP_xi_LP_eff}, \ref{fig:example_3_hist_lambdas_xi_VARMA_1_1_fit} are normalized histograms of the sample and shrunk eigenvalues (bars), and the corresponding eigenvalue densities (lines). Figure \ref{fig:example_3_plot} contains plots of the shrinkage relationship $\xi(\lambda)$ considered in the previous figures, as well the cross-validation-estimated oracle eigenvalues (triangles).}
    \label{fig:example_3}
\end{figure*}

\subsection{Example 1: Gaussian + VAR(1)}

First, we assume a multivariate Gaussian distribution of $Y$, with the underlying true covariance matrix $C$ having two distinct eigenvalues, $1$ and $3$, in 50\% proportions, and with the auto-correlation matrix $A$ given by the exponential decay model (\ref{eq:exp_decay}), with the auto-correlation time $\tau = 3$. This case encompasses the first group of figures \ref{fig:example_1}.

Figures~\ref{fig:example_1_hist_lambdas_xi_oracle_mwcv_iso}, \ref{fig:example_1_hist_lambdas_xi_LP_xi_LP_eff}, \ref{fig:example_1_hist_lambdas_xi_VAR_1_fit} all show the same histogram of the sample eigenvalues $\lambda_{i}$'s (\textit{peach}), as well as the Epanechnikov-kernel-approximated density of eigenvalues $\rho_{E}(\lambda)$ (\textit{orange}), enveloping the histogram. We see how broad this distribution is, completely obscuring behind statistical noise the true eigenvalues $1$ and $3$. We expect that shrinkage will narrow down this histogram, making it more localized about the true eigenvalues.

The first shrinkage approach we choose is the simplest, but very powerful isotonic regression fitted to the cross-validation-estimated oracle eigenvalues (\ref{eq:xi_oracle_mwcv}) (with $K = 10$ and $T_{\textrm{out}} = 50$). Their histogram (\textit{aqua}) and density (\textit{teal}) in figure~\ref{fig:example_1_hist_lambdas_xi_oracle_mwcv_iso} reveal two clear peaks close to $1$ and $3$. The very low Frobenius ratio of 12\% further confirms how effective this shrinkage prescription is.

In figure~\ref{fig:example_1_hist_lambdas_xi_LP_xi_LP_eff} we suppose for a moment that we are ignorant of the fact that there are auto-correlations in $Y$, and we choose to apply the Ledoit-P\'{e}ch\'{e} shrinkage (\ref{eq:LP}) (which recall is valid for the sandwich model with no auto-correlations (\ref{eq:sandwich_1})). The histogram (\textit{pale pink}) and density (\textit{pale purple}) of these $\xi^{\textrm{LP}}_{i}$ is only slightly narrowed as compared to the sample spectrum. The Frobenius ratio of 53\% is large.

One might argue at this point \cite{ref} that the reason why the Ledoit-P\'{e}ch\'{e} formula does not shrink the eigenspectrum enough is that it underestimates the noise level $q = N/T$. Indeed, when auto-correlations with a characteristic time $\tau_{\textrm{eff}}$ are present in the system, the effective number of samples is $T_{\textrm{eff}} = T (1 - \textrm{e}^{- 1 / \tau_{\textrm{eff}}}) \sim T / \tau_{\textrm{eff}}$, smaller than $T$. An interesting exercise would therefore be to find an effective number of samples (by varying $\tau_{\textrm{eff}}$) such that the Ledoit-P\'{e}ch\'{e} estimator performs better in terms of the Frobenius ratio. A caveat here is that it would be illegal to minimize the Frobenius ratio itself; it is, after all, an unknown quantity, dependent on $C$. One can however employ the following trick: fit rather to the scatter points of the cross-validation-estimated oracle eigenvalues (by minimizing the MSE between the two). Indeed, we thus get $T_{\textrm{eff}} = 426$, with the Frobenius ratio $\mathcal{F} = 33\%$. It is significantly lower than for the true-$T$ Ledoit-P\'{e}ch\'{e} result, thus demonstrating the validity of this effective approach. Visually, the histogram (\textit{lilac}) and density (\textit{violet}) in figure~\ref{fig:example_1_hist_lambdas_xi_LP_xi_LP_eff} start slowly developing two peaks in the vicinity of $1$ and $3$.

Finally, apply our VARMA shrinkage formula (\ref{eq:main_3}). Since the data-generating process is known, and in particular $A$ is that of an exp-decay model with $\tau = 3$, one could be tempted to use the same VARMA parameters in the shrinkage formula. But one should immediately object that this information is not available in any real-world scenario. However, just like with the effective $T_{\textrm{eff}}$ above, there is a way to \textit{fit} $\tau$ to the data even without knowing $C$; indeed, to the cross-validation-estimated oracle eigenvalues. In this way, $\tau$ becomes an effective parameter, giving our model greater flexibility than the Ledoit-P\'{e}ch\'{e} case. Figure~\ref{fig:example_1_hist_lambdas_xi_VAR_1_fit} shows the histogram (\textit{mauve}) and density (\textit{maroon}) of the shrunk eigenvalues $\xi_{i}$. The fit here is $\tau = 2.9$ (probably by some numerical accident slightly different from $3$), and we observe how the broad expansion of the sample eigenvalues is efficiently shrunk to two peaks close to $1$ and $3$, the true eigenvalues. The Frobenius ratio is much lower, too, at about 12\%. This number is comparable to the (much simpler) isotonic regression estimator from figure~\ref{fig:example_1_hist_lambdas_xi_oracle_mwcv_iso}.

\subsection{Example 2: Student-t + VMA(1)}

To further simulate a real-world situation, generate now the data matrix $Y$ from the Student-t distribution with $\mu = 5$ degrees of freedom (i.e., with heavy tails), and with the auto-correlation matrix $A$ of the VMA(1) (vector moving average) model with parameters $a_{0} = 0.8$, $a_{1} = 0.5$, cf. (\ref{eq:VARMA}). In other words, the distribution is quite far from the Gaussian, nor the correlations between samples are exponentially decaying. (The model for $C$ stays as before, with two peaks at $1$ and $3$.)

Looking at figure~\ref{fig:example_2} we see that now all the four shrinkage estimators perform quite similarly, with the best isotonic regression at 22\% Frobenius ratio, the LP at 33\%, the effective LP (with fitted $T_{\textrm{eff}} = 811$) at 26\%, while our exp-decay shrinkage (with fitted $\tau = 1.1$) has 24\%. Two peaks are visible in all the cases.

\subsection{Example 3: Student-t + VARMA(1, 1)}

To make the data distribution even more complicated, generate now $Y$ from the Student-t distribution with $3$ degrees of freedom (so, very heavy tails), $C$ with eigenvalues following the inverse-Wishart distribution with $\kappa = 2$, and $A$ that of a VARMA(1, 1) model (\ref{eq:VARMA}) with $a_{0} = 0.8$, $a_{1} = 0.5$, $b_{1} = 0.4$. Recall~\cite{BBP} that for $C$ to have random eigenvalues from the \textit{inverse-Wishart} distribution with a parameter $\kappa$, we first calculate a noise ratio $q_{\textrm{IW}} = 1 / (1 + 2 \kappa)$, and the corresponding number of samples, $T_{\textrm{IW}} = \lfloor N / q_{\textrm{IW}} \rfloor$. Generate an i.i.d. Gaussian random matrix $R$ of shape $N \times T_{\textrm{IW}}$, and form from it a Wishart random matrix, $W = \frac{1}{T_{\textrm{IW}}} R R^{\top}$, i.e., its sample estimator. Finally, $C = (1 - q_{\textrm{IW}}) W^{-1}$.

Figure~\ref{fig:example_3} reveals the same set of graphs as in the previous two examples. There is one difference in the setup, though: Shown in figure~\ref{fig:example_3_hist_lambdas_xi_VARMA_1_1_fit} are the shrunk eigenvalues corresponding not to an exp-decay shrinkage with some optimal $\tau$ fitted to the cross-validation-estimated oracle eigenvalues, as we did before, but a more general three-parameter VARMA(1, 1) shrinkage. We fit $a_{0}$, $a_{1}$, $b_{1}$ by brute-force, simply iterating over a certain grid of values (quite coarse, as this is simply a proof-of-concept), and selecting those for which the MSE distance to the estimated oracle eigenvalues is the smallest. And the main finding from this analysis is that the Frobenius ratio of 27\% of such an optimal fit outperforms the isotonic regression estimator, the latter with 34\%.

In other words, despite severe model mismatch, the mere fact of incorporating auto-correlations into the shrinkage formula vastly improves the quality of eigenvalue cleaning, sometimes even better than the numerical isotonic regression scheme. This bodes well for empirical, real-world applications, where the underlying data-generating processes are genuinely unknown.

\section{Conclusions}

The main result of this paper is an exact formula (\ref{eq:main_1}), (\ref{eq:main_2}), (\ref{eq:main_3}) for the rotationally-invariant estimator (RIE, that is, sharing the eigenbasis with the standard sample estimator (\ref{eq:E}), but with different eigenvalues) (\ref{eq:RIE}) of the true covariance matrix (\ref{eq:C}), which is optimal w.r.t. the mean squared error (\ref{eq:MSE}). The result is crucial in the ``big data'' regime (\ref{eq:big_data}), when the number of correlated entities $N$ is large and of comparable magnitude with the number of collected samples $T$, since then the classical sample estimator contains mostly statistical noise instead of any information about actual correlations. The main novelty of our formula is that it allows for correlations between samples, thus extending the method of Ledoit and P\'{e}ch\'{e}~\cite{LP}. Such auto-correlations effectively decrease the number of samples $T$, thus making the estimation task even harder; a specifically tailored solution such as we present here is therefore of great importance.

We couple our analytical expression with a straightforward numerical scheme based on the kernel estimation method of Ledoit and Wolf. This means a numerically stable and effective pipeline which leads directly from the observed dataset to the cleaned/shrunk eigenvalues. We implement this end-to-end in an open-source Python library, encouraging its use to practitioners of various fields of science and industry where the problem of covariance estimation in the big data regime appears.

We leave for further work the following questions: One clear task is to apply our shrinkage formula in a realistic situation, for instance, a Markowitz-type portfolio optimization, on different markets, hopefully demonstrating that taking into account auto-correlations leads to better performance.

Another question is to consider more general, but still tractable, models of auto-correlations, by which we mean that their $\chi$-transform is analytically accessible, akin to (\ref{eq:chi_exp_decay}). We have already checked that some low-order VARMA models, in particular VAR(1), VMA(1), VARMA(1, 1), VAR(2), VMA(2), belong to this class, and in fact, they are now available in the \texttt{shrinkage} library. In fact, we have used here a VARMA(1, 1) model; it would be very beneficial to devise better methods of fitting parameters of such higher-order models (instead of a brute-force evaluation over a given grid), and interesting to see how they perform, especially compared to the isotonic regression estimator.

We have talked about how EWMA and Maronna estimators naturally fit within our framework of auto-correlations. One task would be to investigate the performance of an EWMA estimator in a situation of non-stationary data, such as on financial markets. On the other hand, the Maronna estimator has a much more involved form, and it would be interesting to see if it can be handled analytically.

This whole work falls within the class of RIEs, in which the key assumption is that we possess no belief about the eigenvectors of $C$. Once such a bias is available, though, for example as a \textit{factor model}, one could attempt to work out an appropriate optimal estimator.

Finally, and more speculatively, one could look at \textit{non-symmetric} correlations, such as between different entities at \textit{different} time moments, say with a time lag of one. Would it be possible to extend the matrix Mar\v{c}enko-Pastur equation to this case, working out a proper generalization of the replica trick and the HCIZ integral?

\begin{acknowledgements}
We thank Artur \'{S}wi\k{e}ch, Vincent W.C. Tan, and Christopher Wells for interesting discussions.
\end{acknowledgements}

\appendix

\section{Scalar Mar\v{c}enko-Pastur equation and free products of rectangular matrices}
\label{app:MP_scalar}

In this appendix we sketch a derivation of the scalar Mar\v{c}enko-Pastur equation (\ref{eq:MP_scalar_1}). An important auxiliary result concerns multiplying \textit{rectangular} random matrices. If $W$ is $N \times T$, while $V$ is $T \times N$, then $W V$ is $N \times N$, and $V W$ is $T \times T$. Since the moments (\ref{eq:moments}) are normalized by the matrix dimension, we find, due to the cyclic property of trace, $m_{V W, k} = \frac{1}{T} \textrm{Tr} \, (V W)^{k} = q \frac{1}{N} \textrm{Tr} \, (W V)^{k} = q m_{W V, k}$. Thus, the scalar moment generating function (\ref{eq:m}) satisfies the same relation, $m_{V W}(z) = q m_{W V}(z)$. A short manipulation leads then to the corresponding relation between the $S$-transforms (\ref{eq:S}) \cite{BJJNPZ},
\begin{equation}\label{eq:S_rect}
    S_{W V}(z) = \frac{q (1 + z)}{1 + q z} S_{V W}(q z) .
\end{equation}
For brevity, denote the prefactor here by $f$, as it won't be important in the following calculation; the important piece is the $q$ in the argument on the right hand side.

Consider now the $S$-transform of the sample estimator in the sandwich model (\ref{eq:sandwich_4}). Since the $S$-transform depends on the matrix through its moments, we have as a consequence of the cyclic property of trace that,
\begin{displaymath}
    S_{E}(z) = S_{\frac{1}{T} X A X^{\top} C}(z) = S_{\frac{1}{T} X A X^{\top}}(z) S_{C}(z) ,
\end{displaymath}
where the second equality comes from the free multiplication law (\ref{eq:mult}). Now, in the first term, we move $X^{\top}$ to the front; this changes the dimension of the matrix, so we need to compensate according to (\ref{eq:S_rect}),
\begin{displaymath}
    S_{\frac{1}{T} X A X^{\top}}(z) = f S_{\frac{1}{T} X^{\top} X A}(q z) = f S_{\frac{1}{T} X^{\top} X}(q z) S_{A}(q z) ,
\end{displaymath}
where again the second equality is the multiplication law. We now again move $X$ to the front, in the first term, obtaining for the last expression the value of $S_{\frac{1}{T} X X^{\top}}(z) S_{A}(q z)$, according to (\ref{eq:S_rect}). The first matrix here is the standard Wishart random matrix (\ref{eq:W}), and so we can write altogether,
\begin{displaymath}
    S_{E}(z) = S_{C}(z) S_{A}(q z) S_{W}(z) ,
\end{displaymath}
with the $q$ in the argument of $S_{A}$ reminiscent of (\ref{eq:S_rect}). The $S$-transform of the Wishart matrix is well-known \cite{JNPWZ}, $S_{W}(z) = 1/(1 + q z)$, which completes our proof of the scalar Mar\v{c}enko-Pastur equation (\ref{eq:MP_scalar_1}), based on the free multiplication law (\ref{eq:mult}).

\end{document}